# Generalised Fluctuation Formula

Debra J. Searles†, Gary Ayton* and Denis J. Evans*

*Research School of Chemistry, Australian National University, Canberra, ACT 0200, Australia
and †Department of Chemistry, University of Queensland, Brisbane, QLD 4072, Australia

**Abstract**. We develop a General Fluctuation Formula for phase variables that are odd under time reversal. Simulations are used to verify the new formula.

## INTRODUCTION

The Fluctuation Theorem (FT) (1-10) gives a simple relationship for the probability of observing trajectory segments with a specific time averaged value of a given phase function, to the probability of observing segments with a negative value: that is, for a phase function $\phi(\Gamma)$, it gives an exact expression for $\Pr(\bar{\phi}(\tau) = A) / \Pr(\bar{\phi}(\tau) = -A)$, where $\bar{\phi}(\tau)$ is the phase function averaged over a time $\tau$. Thus far, FT has only been obtained for a restricted set of phase functions (7) and originally work focussed on the isoenergetic dynamics of a microcanonical ensemble of initial phases $\Gamma$. The only phase function that was considered was the time-averaged rate of entropy production per unit volume, $\bar{\sigma}(\tau)$ which is directly proportional to the dissipative flux and the phase space contraction rate (1-7). This fluctuation formula is of particular interest because it gives an analytic expression of the probability that, for a finite system and for a finite time, the Second Law of thermodynamics will be violated. In addition, FT has been derived from the Sinai-Ruelle-Bowen measure (5,6) and hence provides support for the applicability of this measure to nonequilibrium steady state systems. More recently, FT for other ensembles have been obtained (7). In these cases FT was not expressed in terms of the phase space contraction but only in terms of the dissipative flux and an analytic expression is obtained which is consistent with the Second Law of thermodynamics. In the present paper we obtain General Fluctuation Formulae for any phase variable that is odd under time-reversal symmetry.

## DERIVATION OF THE GENERALISED FLUCTUATION THEOREM (GFT)

Consider an N-particle system with coordinates and peculiar momenta, $\{\mathbf{q}_1, \mathbf{q}_2, .. \mathbf{q}_N, \mathbf{p}_1, ..\mathbf{p}_N\} \equiv (\mathbf{q}, \mathbf{p}) \equiv \Gamma$. The internal energy of the system is $H_0 \equiv \sum_{i=1}^{N} p_i^2 / 2m + \Phi(\mathbf{q}) = K + \Phi$ where $\Phi(\mathbf{q})$ is the interparticle potential energy, which is a function of the coordinates, **q**, of all the particles and K is the total peculiar kinetic energy. In the presence of an external field $F_e$, the thermostatted equations of motion are taken to be,

$$\dot{\mathbf{q}}_i = \mathbf{p}_i / m + \mathbf{C}_i(\Gamma)F_e$$
$$\dot{\mathbf{p}}_i = \mathbf{F}_i(\mathbf{q}) + \mathbf{D}_i(\Gamma)F_e - \alpha(\Gamma)\mathbf{p}_i \qquad (1)$$

where $\mathbf{F}_i(\mathbf{q}) = -\partial \Phi(\mathbf{q}) / \partial \mathbf{q}_i$ and $\alpha$ is the thermostat multiplier which in this case is applied to the peculiar momenta, and $\mathbf{C}_i$ and $\mathbf{D}_i$ represent the coupling of the system to the field. The dissipative flux J, is defined in terms of the adiabatic time derivative of the internal energy, $\dot{H}_0^{ad} \equiv -J(\Gamma)VF_e$.

Let $\phi(\Gamma)$ be an arbitrary phase function and define the time average

$$\overline{\phi}_i(\tau) = \frac{1}{\tau}\int_0^\tau ds\, \phi(\Gamma_i(s)). \qquad (2)$$

At t = 0 the phase space volume occupied by a contiguous bundle of trajectories for which $\{\Gamma_i | A < \overline{\phi}_i(\tau) < A + \delta A\}$ is given by $\delta V(\Gamma(0), 0)$ and at time $\tau$ these phase points will occupy a volume $\delta V(\Gamma(\tau), \tau) = \delta V(\Gamma(0), 0)e^{\overline{\Lambda}(\tau)\tau}$ where $\overline{\Lambda}(\tau)$ is the time-averaged phase space compression factor along these trajectories. We denote $\overline{\phi}(\tau) = <\overline{\phi}_i(\tau)>_{\{\Gamma\}}$, that is the average value of $\overline{\phi}_i(\tau)$ over the set of contiguous trajectories, $\{\Gamma_i\}$.

If the dynamics is reversible, there will be a contiguous set of initial phases $\{\Gamma_i^*(0)\}$, given by $\Gamma_i^*(0) = M^T(\Gamma_i(\tau))$, that will occupy a volume $\delta V(\Gamma^*(0), 0) = \delta V(\Gamma(\tau), \tau) = \delta V(\Gamma(0), 0)e^{\overline{\Lambda}(\tau)\tau}$ along which the time-averaged value of the phase function is $\overline{\phi}_{i*}(\tau) = M^T(\overline{\phi}_i(\tau))$. For any $\phi_i(\Gamma)$ that is odd under time reversal, $\overline{\phi}_{i*}(\tau) = -\overline{\phi}_i(\tau)$.

The probability ratio of observing trajectories originating in an initial phase volume and its conjugate phase volume will be related to the initial phase space distribution function and the size of the volume elements:

$$\frac{\Pr(\delta V(\Gamma(0), 0))}{\Pr(\delta V(\Gamma^*(0), 0))} = \frac{f(\Gamma(0), 0)\delta V(\Gamma(0), 0)}{f(\Gamma^*(0), 0)\delta V(\Gamma^*(0), 0)}$$

$$= \frac{f(\Gamma(0), 0)}{f(\Gamma(\tau), 0)} e^{-\overline{\Lambda}(\tau)\tau} \qquad (3)$$

$$= e^{\overline{\text{ß}}(\tau)\tau}$$

where

$$\overline{\text{ß}}(\tau)\tau \equiv \ln\left(\frac{f(\Gamma(0), 0)}{f(\Gamma(\tau), 0)}\right) - \int_0^\tau \Lambda(\Gamma(s))ds, \qquad (4)$$

then $\ln(f(\Gamma(0),0)/f(\Gamma(\tau),0)) = 0,$ and $\overline{\text{ß}}(\tau)\tau = -\overline{\Lambda}(\Gamma(\tau))\tau = -\beta\overline{J}(\tau)VF_e\tau$ where $\beta = dN/2K$ and d is the Cartesian dimensionality. If the initial ensemble is canonical, then $\overline{\text{ß}}(\tau)\tau = \beta(\Phi(\tau) - \Phi(0)) - \overline{\Lambda}(\tau)\tau = -\beta\overline{J}(\tau)VF_e\tau$.

It is possible that there are non-contiguous bundles of trajectories for which $\{\Gamma_i | A < \overline{\phi}_{\tau,i} < A + \delta A\}$, and since these bundles may have different values $\overline{\text{ß}}(\tau)$ the probability ratio (Eqn (3)) may differ for each bundle. The probability of observing a trajectory for which $A < \overline{\phi}(\tau) < A + \delta A$, is obtained by summing over the probabilities of observing these m=1,M non-contiguous volume elements, $\delta V_m(\Gamma(0),0)$. If the phase function is odd under time reversal symmetry,[1] then the ratio of the probability of observing trajectories for which $A < \overline{\phi}(\tau) < A + \delta A$ to the probability of observing conjugate trajectories, for which $-A < \overline{\phi}(\tau) < -A + \delta A$ is,

$$\frac{\Pr(\overline{\phi}(\tau) = A)}{\Pr(\overline{\phi}(\tau) = -A)} = \frac{\sum_{m=1}^{M} \Pr(\delta V_m(\Gamma(0),0))}{\sum_{m=1}^{M} \Pr(\delta V_m(\Gamma^*(0),0))}$$

$$= \frac{\sum_{m=1}^{M} \Pr(\delta V_m(\Gamma(0),0))}{\sum_{m=1}^{M} \Pr(\delta V_m(\Gamma(0),0)) e^{-\overline{\text{ß}}(\tau)\tau}} \quad (5)$$

$$= \left\langle e^{-\overline{\text{ß}}(\tau)\tau} \right\rangle^{-1}_{\overline{\phi}(\tau)=A}$$

where the notation $\langle...\rangle_{\overline{\phi}_i(\tau)=A}$ refers to the ensemble average over contiguous trajectory bundles for which $\overline{\phi}(\tau) = A$. Equation (5) gives the ratio of the measure of those phase space trajectories for which $\overline{\phi}(\tau) = A$ to the measure of those trajectories for which $\overline{\phi}(\tau) = -A$. This is the Generalised Fluctuation Formula for any phase variable $\overline{\phi}(\tau)$ that is odd under time reversal. The actual form of $\overline{\phi}(\tau)$ is quite arbitrary.

For isoenergetic dynamics initiated from a microcanonical ensemble (7),

$$\frac{\Pr(\overline{\phi}(\tau) = A)}{\Pr(\overline{\phi}(\tau) = -A)} = \left\langle e^{\overline{\Lambda}(\tau)\tau} \right\rangle^{-1}_{\overline{\phi}(\tau)=A} = \left\langle e^{VF_e\beta\overline{J}(\tau)\tau} \right\rangle^{-1}_{\overline{\phi}(\tau)=A} \quad (6)$$

likewise for isokinetic or Nosé-Hoover dynamics initiated from a canonical ensemble (7),

---

[1] If the phase variable is even, the we obtain the trivial relationship

$$\left\langle e^{-\overline{\text{ß}}(\tau)\tau} \right\rangle^{-1}_{\overline{\phi}(\tau)=A} = \frac{\Pr(\overline{\phi}(\tau) = A)}{\Pr(\overline{\phi}(\tau) = A)} = 1$$

$$\frac{\Pr(\bar{\phi}(\tau) = A)}{\Pr(\bar{\phi}(\tau) = -A)} = \left\langle e^{VF_e\beta\bar{J}(\tau)\tau} \right\rangle^{-1}_{\bar{\phi}(\tau)=A}. \qquad (7)$$

Formulas for other ergodically consistent ensembles can be obtained in a similar manner (7).

## CORRELATIONS BETWEEN TIME AVERAGES AND THE ENTROPY PRODUCTION

Equation (5) is the generalised fluctuation formula for properties, $\bar{\phi}(\tau)$ which are odd under time reversal mapping. The time averaged entropy production rate per unit volume, $\bar{\sigma}(\tau)$ is also odd under time reversal symmetry, however $\bar{\phi}(\tau)$ does not *necessarily* change sign with $\bar{\sigma}(\tau)$. For example, the kinetic component of the shear stress, $\bar{P}^K_{xy}(\tau)$ is odd under time reversal, but does not necessarily have the same sign as $\bar{\sigma}(\tau)$. One has to include the possibility that when $\bar{\phi}(\tau)<0$, $\bar{\sigma}(\tau) > 0$ and when $\bar{\phi}(\tau) >0$, $\bar{\sigma}(\tau) < 0$. The probability ratio in Equation (5) can be written as

$$\frac{\Pr(\bar{\phi}(\tau) = A)}{\Pr(\bar{\phi}(\tau) = -A)} = \frac{\Pr(\sigma > 0)\Pr(\bar{\phi}(\tau) = A;\sigma > 0) + \Pr(\sigma < 0)\Pr(\bar{\phi}(\tau) = A;\sigma < 0)}{\Pr(\sigma > 0)\Pr(\bar{\phi}(\tau) = -A;\sigma > 0) + \Pr(\sigma < 0)\Pr(\bar{\phi}(\tau) = -A;\sigma < 0)}. \qquad (8)$$

We can consider Equation (8) in more detail. The probability of observing A, $\Pr(\bar{\phi}(\tau) = A)$ has been decomposed into two terms. The first "direct" term is the product of $\Pr(\bar{\sigma}(\tau) > 0)$, the probability of observing a trajectory with positive entropy production, and $\Pr(\bar{\phi}(\tau) = A;\bar{\sigma}(\tau) > 0)$, the probability of observing $\bar{\phi}(\tau) = A$ on a segment with $\bar{\sigma}(\tau) > 0$. The second term can be considered a "cross term" and accounts for cases when $\bar{\sigma}(\tau) < 0$ *and* $\bar{\phi}(\tau) >0$. Likewise, $\Pr(\bar{\phi}(\tau) = -A)$ is expressed in terms of a direct term $\Pr(\bar{\sigma}(\tau) < 0)\Pr(\bar{\phi}(\tau) = -A;\bar{\sigma}(\tau) < 0)$ which accounts for trajectories with $\bar{\phi}(\tau)<0$, $\bar{\sigma}(\tau) < 0$, and a "cross term" $\Pr(\bar{\sigma}(\tau) > 0)\Pr(\bar{\phi}(\tau) = -A;\bar{\sigma}(\tau) > 0)$. We can describe the degree of *correlation* between $\bar{\phi}(\tau)$ and $\bar{\sigma}(\tau)$ by considering the direct and cross term contributions in the probability ratio. Choices of $\bar{\phi}(\tau)$ that are highly correlated with $\bar{\sigma}(\tau)$ imply that the actual *value* of $\bar{\sigma}(\tau)$ strongly depends on A; in the limit when $\bar{\phi}(\tau)=\bar{\sigma}(\tau)$, the cross terms are exactly zero and

$$\frac{\Pr(\bar{\phi}(\tau) = A)}{\Pr(\bar{\phi}(\tau) = -A)} = \frac{\Pr(\bar{\sigma}(\tau) > 0)\Pr(\bar{\phi}(\tau) = A;\bar{\sigma}(\tau) > 0)}{\Pr(\bar{\sigma}(\tau) < 0)\Pr(\bar{\phi}(\tau) = -A;\bar{\sigma}(\tau) < 0)}. \qquad (9)$$
$$= \exp[-\bar{\Lambda}(\tau)\tau]$$

In cases where $\bar{\phi}(\tau)$ is uncorrelated with $\bar{\sigma}(\tau)$, $\Pr(\bar{\phi}(\tau) = A;\bar{\sigma}(\tau) > 0) = \Pr(\bar{\phi}(\tau) = A;\bar{\sigma}(\tau) < 0)$ and $\Pr(\bar{\phi}(\tau) = -A;\bar{\sigma}(\tau) > 0) = \Pr(\bar{\phi}(\tau) = -A;\bar{\sigma}(\tau) < 0)$, thus

$$\frac{\Pr(\bar{\phi}(\tau) = A)}{\Pr(\bar{\phi}(\tau) = -A)} = \left\langle e^{\overline{\Lambda}(\tau)\tau} \right\rangle^{-1}_{\bar{\phi}(\tau)=A} = 1, \tag{10}$$

where $\left\langle e^{\overline{\Lambda}(\tau)\tau} \right\rangle^{-1}_{\bar{\phi}(\tau)=A}$ is the Kawasaki normalization factor (3).

## RESULTS

To demonstrate the validity of GFT we have selected five possible phase functions $\bar{\phi}(\tau)$ that are odd under time reversal. Different components related to the shear stress, a non-linear function of the shear stress, and a test function designed to be completely unrelated to the entropy production will be considered. Numerical testing of GFT was done with NEMD simulations of N=32 WCA particles in two Cartesian directions under isoenergetic shear flow using the SLLOD equations of motion for planar Couette flow (11). The equations of motion are

$$\begin{aligned}\dot{\mathbf{q}}_i &= \mathbf{p}_i / m + \mathbf{i}\gamma y_i \\ \dot{\mathbf{p}}_i &= \mathbf{F}_i - \mathbf{i}\gamma p_{yi} - \alpha \mathbf{p}_i\end{aligned} \tag{11}$$

and $\gamma$ is the applied strain rate and $\alpha$ is the Gaussian isoenergetic ergostat multiplier. In all simulations the number density was n=0.8, the internal energy per particle was E/N=1.056, $\gamma$=0.2, and time averaging interval was $\tau$=0.08. The isoenergetic SLLOD equations of motion are reversible. If we choose the strain rate to be invariant under the time reversal mapping, the dissipative flux and the entropy production are odd and the reverse trajectory is generated by carrying out the Kawasaki mapping of a point in phase space at time t, $M^K(x, y, p_x, p_y) = (x, -y, -p_x, p_y)$ and then propagating forward in time (2,11).

As previously discussed, general phase functions that are antisymmetric with respect to time reversal need not be correlated with the entropy production. To demonstrate this, we consider four functions with varying degrees of entropy production correlation, beginning with a highly correlated function ($\overline{\beta P}_{xy}(\tau)$ itself), then strongly correlated (the potential component of $\overline{\beta P}_{xy}(\tau)$, $\overline{\sum_{i=1}^{N} p_{xi}^3(\tau)}$), weakly correlated (the kinetic component of $\overline{\beta P}_{xy}(\tau)$, $\overline{\beta P}^K_{xy}(\tau)$) and finally an uncorrelated[2] phase function ($\overline{\sum_{i=1}^{N} p_{xi}^3(\tau)}$, where $p_{xi}$ is the x-component of the peculiar momenta for the i$^{th}$ particle).

The GFT for $\bar{\phi}(\tau) = \overline{\beta P}^K_{xy}(\tau)$ can be written as

___

[2] In practice, it is difficult to choose a "perfectly uncorrelated" phase function, therefore strickly speaking, by "uncorrelated" we mean that there is little or no measureable correlation between the phase function, and the time averaged entropy production.

$$\frac{\Pr(\overline{\beta P}_{xy}^{K}(\tau) = A)}{\Pr(\overline{\beta P}_{xy}^{K}(\tau) = -A)} = \left\langle e^{\overline{\beta P}_{xy}(\tau)VF_e\tau} \right\rangle^{-1}_{\overline{\beta P}_{xy}^{K}(\tau)=A}$$

$$= \left\langle e^{\left[\overline{\beta P}_{xy}^{K}(\tau)+\overline{\beta P}_{xy}^{P}(\tau)\right]VF_e\tau} \right\rangle^{-1}_{\overline{\beta P}_{xy}^{K}(\tau)=A} \qquad (12)$$

$$= \frac{1}{e^{AVF_e\tau}\left\langle e^{\overline{\beta P}_{xy}^{P}(\tau)VF_e\tau} \right\rangle_{\overline{\beta P}_{xy}^{K}(\tau)=A}}$$

where the correlation between $\overline{\beta P}_{xy}^{P}(\tau)$ and $\overline{\beta P}_{xy}^{K}(\tau)$ determines the "degree" of correlation between $\overline{\beta P}_{xy}^{K}(\tau)$ and $\overline{\sigma}(\tau)$. A similar formula can be constructed for $\Pr(\overline{\beta P}_{xy}^{P}(\tau) = A)/\Pr(\overline{\beta P}_{xy}^{P}(\tau) = -A)$.

In Figure 1 we show that $\Pr(\overline{\beta P}_{xy}^{P}(\tau) = A)/\Pr(\overline{\beta P}_{xy}^{P}(\tau) = -A)$ agrees with $\left\langle e^{\overline{\beta P}_{xy}(\tau)VF_e\tau} \right\rangle^{-1}_{\overline{\beta P}_{xy}^{P}(\tau)=A}$, as predicted by the GFT (Eqn (5)).

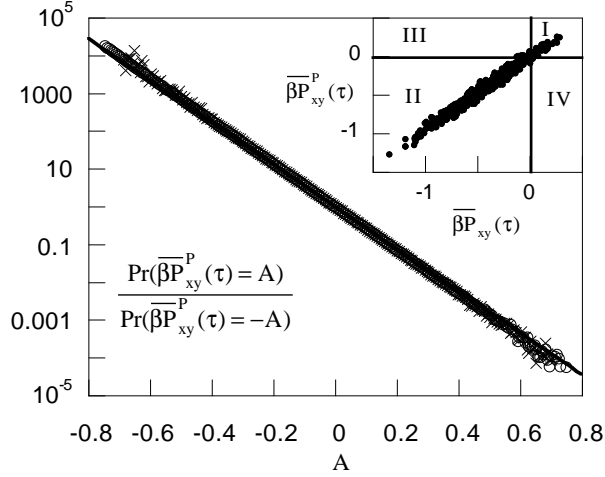

**FIGURE 1.** A plot of $\Pr(\overline{\beta P}_{xy}^{P}(\tau) = A) / \Pr(\overline{\beta P}_{xy}^{P}(\tau) = -A)$ versus A for a system of 32 WCA particles undergoing shear flow. The crosses correspond to the directly determined probability ratio, and the open circles correspond to $\left\langle e^{\overline{\beta P}_{xy}VF_e\tau} \right\rangle^{-1}_{\overline{\beta P}_{xy}^{P}(\tau)=A}$. The inset shows the relationship between $\overline{\beta P}_{xy}(\tau)$ and $\overline{\beta P}_{xy}^{P}(\tau)$ for a randomly selected set of transient trajectories, giving an indication of the relative direct and cross contributions to the probability ratio. The line shows the results expected if $\overline{\beta P}_{xy}(\tau)=\overline{\beta P}_{xy}^{P}(\tau)$ (perfect correlation).

The observed linear slope is also very close to the slope found for FT using $\overline{P}_{xy}(\tau) = A$ (shown as the solid line), indicating that the potential contribution of the shear stress is highly correlated to the entropy production rate. This is hardly surprising since at this thermodynamic state point 90% of the shear stress comes from the potential rather than kinetic, contributions. In the inset we show $\overline{\beta P}_{xy}(\tau)$ and $\overline{\beta P}^P_{xy}(\tau)$ for a randomly selected set of time averaged transient trajectories. Those points within regions I and II correspond to "direct" contributions to the probability ratio, that is those trajectories with $A > 0$, $\overline{\beta P}^P_{xy}(\tau) = A$, $\overline{\sigma}(\tau) > 0$, and $\overline{\beta P}^P_{xy}(\tau) = -A$, $\overline{\sigma}(\tau) < 0$ respectively. We see that in this highly correlated example, almost all the trajectories reside in these regions, whereas the "cross" contributions, $\overline{\beta P}^P_{xy}(\tau) = A$, $\overline{\sigma}(\tau) < 0$ and $\overline{\beta P}^P_{xy}(\tau) = -A$, $\overline{\sigma}(\tau) > 0$ are near zero.

Figure 2 shows a similar plot for $\overline{\beta P}^K_{xy}(\tau)$. Again GFT (Eqn (5)) is verified although the scatter is much larger. If one incorrectly assumed that the constrained average, $\left\langle e^{\overline{[\beta P]}_{xy}(\tau) V F_e \tau} \right\rangle_{\overline{\beta P}^K_{xy}(\tau) = A}$, could be replaced by $\left\langle e^{\overline{[\beta P]}_{xy}(\tau) V F_e \tau} \right\rangle_{\overline{\beta P}_{xy}(\tau) = A} = e^{A V F_e \tau}$ the result would be that given by the line. Clearly, this is incorrect. An examination of a similar randomly selected set of transient trajectories as shown in the inset indicates that a large proportion of the trajectories originate from the "cross" contributions of region II where $\overline{\beta P}^P_{xy}(\tau) = A$, $\overline{\sigma}(\tau) < 0$, clearly demonstrating that although the kinetic contribution to the shear stress is odd under time reversal, the entropy production observed on a trajectory segment is only weakly correlated with the sign of the time averaged kinetic shear stress. In fact, quite frequently, thermodynamically favorable trajectories $\overline{\sigma}(\tau) > 0$ have a kinetic contribution to the shear stress that has the opposite sign to that required to make $\overline{\sigma}(\tau) > 0$. The largest contribution to $\Pr(\overline{\beta P^K_{xy}}(\tau)) > 0$ actually occurs from region III.

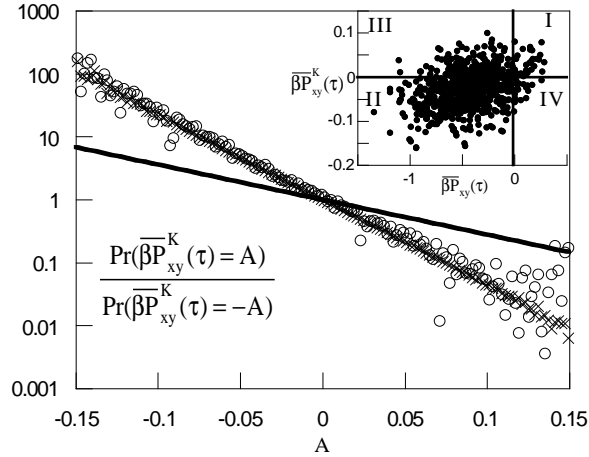

**FIGURE 2**. A plot of $\Pr(\overline{\beta P}^K_{xy}(\tau) = A) / \Pr(\overline{\beta P}^K_{xy}(\tau) = -A)$ versus A for a system of 32 WCA particles undergoing shear flow. The crosses correspond to the directly determined probability ratio, and the open circles correspond to $\left\langle e^{\overline{\beta P}_{xy} V \beta F_e \tau} \right\rangle^{-1}_{\overline{\beta P}^K_{xy}(\tau) = A}$. The inset shows the relationship between $\overline{\beta P}_{xy}(\tau)$ and $\overline{\beta P}^K_{xy}(\tau)$ for a randomly selected set of transient trajectories, giving an indication of the relative direct and cross contributions to the probability ratio. The line shows the results expected if $\overline{\beta P}_{xy}(\tau) = \overline{\beta P}^K_{xy}(\tau)$ (perfect correlation).

In Figure 3 we examine GFT for the "uncorrelated" phase function $\overline{\phi}(\tau) = \overline{\sum_{i=1}^{N} p_{xi}^3(\tau)}$ As predicted, a phase function that is uncorrelated with the entropy production will have $\Pr(\overline{\phi}(\tau) = A) / \Pr(\overline{\phi}(\tau) = -A) = \left\langle e^{\overline{A}(\tau)\tau} \right\rangle^{-1}_{\overline{\phi}(\tau)=A} = 1 \quad \forall A$. Clearly for the values of A tested, GFT holds, and an examination of randomly selected trajectories as shown in the inset confirms that most trajectories reside in regions II and III.

We also consider two functions whose behaviour is expected to be quite different to that of $\overline{\sigma}(\tau)$: $\overline{\phi}(\tau) = \overline{P^3_{xy}}(\tau) = \frac{1}{\tau}\int_0^\tau ds P^3_{xy}(s)$ and the phase function $\phi(\tau) = \left(\overline{\beta P_{xy}}(\tau)\right)^3 = \left(\frac{1}{\tau}\int_0^\tau ds \beta P_{xy}(s)\right)^3$. These functions may be considered as nonlinear functions of the instantaneous entropy production.

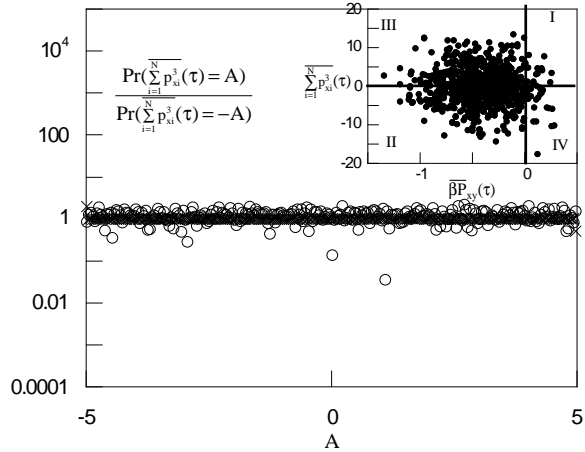

**FIGURE 3.** A plot of $\Pr(\overline{\sum_{i=1}^{N} p_{xi}^3}(\tau) = A) / \Pr(\overline{\sum_{i=1}^{N} p_{xi}^3}(\tau) = -A)$ versus A for a system of 32 WCA particles undergoing shear flow. The crosses correspond to the directly determined probability ratio, and the open circles correspond to $\left\langle e^{\overline{\beta P_{xy} V F_e \tau}} \right\rangle^{-1}_{\overline{\sum_{i=1}^{N}(p_{xi}^3)}(\tau)=A}$. The inset shows the relationship between $\overline{\beta P_{xy}}(\tau)$ and $\overline{\sum_{i=1}^{N} p_{xi}^3}(\tau)$ for a randomly selected set of transient trajectories, giving an indication of the relative direct and cross contributions to the probability ratio.

Equation (3) is tested for these functions in Figures 4 and 5, respectively and the relation is verified by the numerical data. In contrast to the previous FT (1-10), the probability ratio is not an exponential function of the phase function considered.

For $\phi(\tau) = \left(\overline{\beta P}_{xy}(\tau)\right)^3$ it is possible to simplify Equation (3) further, obtaining,

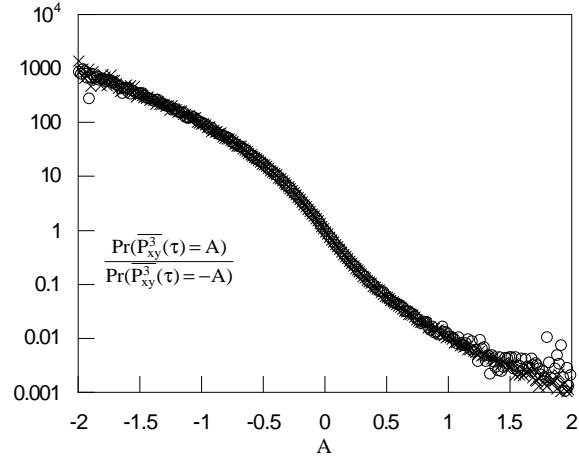

**FIGURE 4.** A plot of $\Pr((\overline{\beta P}_{xy}^3(\tau)) = A) / \Pr((\overline{\beta P}_{xy}^3(\tau)) = -A)$ versus A for a system of 32 WCA particles undergoing shear flow. The crosses correspond to the directly determined probability ratio, and the open circles correspond to $\left\langle e^{\overline{\beta P}_{xy} V F_e \tau} \right\rangle^{-1}_{(\overline{\beta P}_{xy}^3(\tau))=A}$.

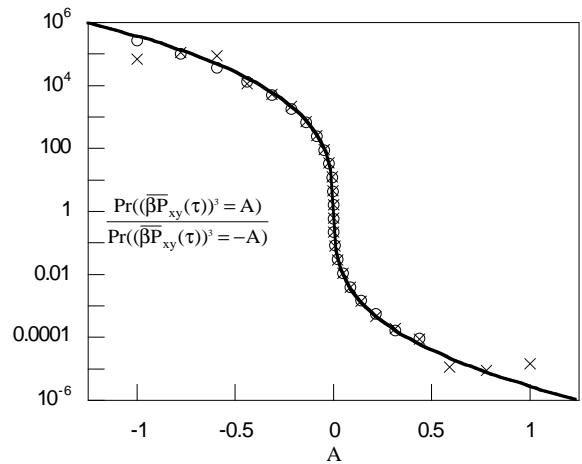

**FIGURE 5.** A plot of $\Pr((\overline{\beta P}_{xy}(\tau))^3 = A) / \Pr((\overline{\beta P}_{xy}(\tau))^3 = -A)$ versus A for a system of 32 WCA particles undergoing shear flow. The crosses correspond to the directly determined probability ratio, and the open circles correspond to $\left\langle e^{\overline{\beta P}_{xy} V F_e \tau} \right\rangle^{-1}_{(\overline{\beta P}_{xy}(\tau))^3=A}$. The line shows the behaviour predicted for the probability ratio by Equation (13).

$$\frac{p((\overline{\beta P}_{xy}(\tau))^3 = A)}{p((\overline{\beta P}_{xy}(\tau))^3 = -A)} = \left\langle \exp(\overline{\beta P}_{xy}(\tau)V\gamma\tau) \right\rangle^{-1}_{(\overline{\beta P}_{xy}(\tau))^3 = A}$$

$$= \left\langle \exp(A^{1/3}V\gamma\tau) \right\rangle^{-1} \quad (13)$$

$$= \exp(-A^{1/3}V\gamma\tau)$$

The curve shown on Figure 5 is the behaviour expected if Equation (13) is true, and the figure indicates that the numerical data is well determined by this relation. This function is not exponential in A, but is exponential in $A^{1/3}$, and this behaviour will persist for all finite $\tau$.

We have obtained a Generalised Fluctuation Formula which is applicable to any phase function that is odd under a time-reversal mapping. The numerical results verify our prediction. Since the formula is quite general it may readily be extended to such problems as those considering a phase variable determined by a restricted set of particles, which results to a local formula (12) or to the average value of a variable over part of a trajectory segment, an approach that may be taken to obtain a steady state fluctuation formula (7,13). Finally, we note that these formulae can also be applied asymptotically to steady state trajectory segments rather than the transient segments considered here.

## ACKNOWLEDGEMENTS

We would like to thank the Australian Research Council for the support of this project. Helpful discussions and comments from Professor E. G. D. Cohen are also gratefully acknowledged.